\documentclass[12pt]{article}
\begin{document}
\baselineskip=16pt

\title
{\begin{Huge}
External Energy Paradigm\\
For Black Holes\\
\end{Huge}
\vspace{0.1in}  }
\author{Yuan K. Ha\\ Department of Physics, Temple University\\
 Philadelphia, Pennsylvania 19122 U.S.A. \\
 yuanha@temple.edu \\    \vspace{.1in}  }
\date{October 1, 2018}
\maketitle
\vspace{.1in}
\begin{abstract}
\noindent
A new paradigm for black holes is introduced. It is known as the External Energy Paradigm. The paradigm asserts that
all energies of a black hole are external quantities; they are absent inside the horizon. These energies include
constituent mass, gravitational energy, electrostatic energy, rotational energy, heat energy, etc. As a result,
quantum particles with charges and spins cannot exist inside the black hole. 
To validate the conclusion, we derive the moment of inertia of a Schwarzschild black hole and find that it is exactly equal to $mass \times (Schwarzschild \,\,\,  radius)^{2}$, indicating that all mass of the black hole is located at the horizon. This remarkable result can resolve
several long-standing paradoxes in black hole theory; such as why entropy is proportional to area and not to volume,
the singularity problem, the information loss problem, and the perplexing firewall controversy.\\

\vspace{.3in}
\noindent
{\em Keywords}: Black holes; horizon mass; Kerr metric; moment of inertia.\\
\end{abstract}

\newpage
A major revision on the structure of the black hole is achieved by investigating the quasi-local energy distribution and 
the rotational properties of the black hole in an exact analytic approach. For over 50 years, the black hole has been considered to be a structureless object with a point singularity shrouded within a horizon. The horizon is a mathematical surface devoid of physical content. It is a definition such that light at which location cannot escape to infinity.
External particles can cross the horizon without catastrophe and they all end up inevitably at the singularity [1].\\ 

But singularity is an illness in physics, not a virtue. A point mass has an infinite negative potential energy which 
would miraculously result in a finite positive mass for any black hole to a distant observer. The singularity is nevertheless 
accepted as an irrefutable outcome for the black hole based on the belief that no known mechanism can prevent matter from collapsing as it reaches the horizon. The Finkelstein-Kruskal coordinate transformation [2,3] on the Schwarzschild spacetime is often cited to justify this point of view.\\

We intend, on the contrary, to establish the black hole with a structure like any other macroscopic object in astrophysics. 
Our definition of a black hole is a massive body with a Schwarzschild radius from which light cannot escape to infinity.
The horizon of a Schwarzschild black hole is replaced by a real physical surface where all mass of the black hole is 
located. It is a `brick-and-mortar' realization of the fascinating Membrane Paradigm [4]. The interior of the black hole is hollow and there is no singularity or gravity present. Spacetime is ordinary and flat. Quantum particles cannot exist inside the black hole because they are forbidden by their very possession of charges and spins. This is the remarkable result of the Horizon Mass Theorem [5]. It is a generalization of the quasi-local energy approach first applied to the static black hole [6,7]. The Horizon Mass Theorem states that: {\em For all black holes; neutral, charged or rotating, the mass observed at the event horizon is always twice the irreducible mass observed at infinity.} The irreducible mass does not contain electrostatic or rotational energy, leading inescapably to the conclusion that
all energies of a black hole are external quantities. This becomes the External Energy Paradigm [8], according to which 
{\em energy and matter cannot exist inside the black hole}. Due to the extreme density of the physical surface, external 
particles cannot cross the horizon and must stay outside or at the surface.\\

The discovery of gravitational waves GW150914 by LIGO confirmed the existence of black holes [9] and unexpectedly 
points out the significance of the concept of the horizon mass. The mass of the black hole observed at the event horizon is crucial for accounting the redshifts of the gravitational waves during their escape in black hole merging [10]. A higher mass at the event horizon and its neighborhood is compulsory in order to overcome the huge negative gravitational potential energy surrounding the black hole. No gravitational wave emission is possible if the horizon mass is the same as the asymptotic mass. The horizon mass further provides an analytical way of determining the mass of the core of the collapsing stars which formed the initial black holes before merging. This in turn allows a quick way to estimate 
the mass of the progenitor stars without having to resort to exhaustive numerical simulations as is excellently done 
in Ref.[11].\\

For this investigation, we shall refer to the Kerr metric [12] for a rotating mass in general relativity 
in subsequent analysis. 
A rotational length parameter, here called alpha, is defined for the angular momentum $J$ and the mass $M$ as 
$ \alpha = J/Mc $ . 
A dimensionless spin parameter $a$ is also used in LIGO analysis of the merging black holes. It is defined as
the angular momentum $J$ compared to the maximum possible angular momentum $J_{max}$ 
for the same mass $M$ 	, i.e.
\begin{equation}
a = \frac{J}{J_{max}} = \frac{J}{ (GM^{2}/c) } = \frac{Jc}{GM^{2}} ,  \hspace{0.5in}  0 \leq a \leq 1 . 
\end{equation} 
The horizon radius is then given by
\begin{equation}  
r_{+} = \frac{GM}{c^{2}} \left[ 1 + \sqrt { 1 - a^{2} } \right] ,
\end{equation}
and the irreducible mass by [13]
\begin{equation}
M_{irr} = \left [ \frac{M^{2}}{2} + \frac{M^{2}}{2} \sqrt{ 1 - a^{2} } \right]^{1/2} .
\end{equation}
The horizon mass is simply
\begin{equation}
M(r_{+}) = 2M_{irr}.
\end{equation}
The difference between the Kerr mass and the irreducible mass is the rotational mass of the black hole, 
$ M - M_{irr} = M_{rot} $ .\\

There are profound implications of the irreducible mass of a general black hole.
Since all electric field lines terminate at electric charges and electrostatic energy is external, this implies
that electric particles cannot exist inside a black hole; they can only stay at the surface. Similarly, rotational 
energy is external, therefore any particle with angular momentum cannot exist inside the black hole and must also
stay outside. According to S. Tomonaga: {\em Nature was not satisfied by a simple point charge
but required a charge with spin} [14]. All matter particles, quarks and leptons, are fermions with both spins and charges; 
therefore constituent matter cannot exist inside the black hole. The External Energy Paradigm states that all energies of a black hole are external quantities. These energies include constituent mass, gravitational energy, electrostatic energy, rotational energy, heat energy, etc.\\

There are several mathematical reasons why energy is forbidden in a classical black hole. The first is that space and time are interchanged inside a black hole as seen by an external observer, i.e.  $ r \rightarrow t $, $ t \rightarrow r $ . 
Since energy dimension is defined as $ [M][L]^{2}[T]^{-2} $ 
in external spacetime, it becomes $ [M][T]^{2}[L]^{-2} $ in the interior and this dimension is unknown in physics.\\
 
A second reason is elementary particles exist only in flat spacetime in quantum field theory. Fermions belong to spinor representations
of the Lorentz group $ SO(3,1) $. There is no spinor representation of the group of general coordinate transformations
 $ GL(4,R) $ in general relativity. At most fermions exist in the tangent space when the curvature is small. 
If the tangent space is too small, then fermions cannot exist in that region of spacetime.\\
 
A third reason is that quantum particles are in reality extended particles with a size given by the Compton wavelength
$ \lambda = h/mc $. They are called point-like particles only because they are described by local interaction terms
at the same point in the Lagrangian in quantum field theory. Despite a hundred years of general relativity, it is still
not known exactly how {\em quantum} particles with spin helicities behave in Schwarzschild spacetime, or whether they always follow the geodesics of classical point masses. All particles in Nature are quantum particles. The conclusion that 
fermions can cross the horizon based on the Kruskal diagram is questionable.\\

A fourth reason concerns the property of an electrically charged black hole. A Schwarzschild black hole becomes 
a Reissner-Nordstr\"{o}m black hole even with the addition of a single electron. A consistent solution in this case
requires the electron to stay at the surface of the black hole since its charge cannot be separated from its mass.
This is one of the fascinating aspects of the Membrane Paradigm. The electron therefore cannot exist inside a Schwarzschild black hole.\\

However, the most compelling reason why quantum particles cannot exist in a black hole is a simple physical one. We can
show that the Schwarzschild black hole has a hollow shell structure with all mass located at the surface. The surface
becomes the horizon and the extreme density of the surface prevents external particles from crossing the horizon.\\

We demonstrate our result with a concise and exact proof from Kerr rotation.  The angular velocity at the event horizon
from the Kerr metric is known to be [15] 
\begin{equation}
\Omega_{+} = \frac{ \alpha c }{ ( r_{+}^{2} + \alpha^{2} ) } .
\end{equation}
An exact relation written explicitly in terms of the angular momentum $J$ and the mass $M$  becomes
\begin{equation}
\Omega(J) =     \frac{ \displaystyle \frac{J}{M} }  { \displaystyle \frac{2G^{2}M^{2}}{c^{4} } 
                \left[ 1 + \sqrt { 1 -  \displaystyle \frac{J^{2}c^{2} }{G^{2}M^{4} } }
                \right ]  } .                                                  
\end{equation}
We may express the angular momentum in terms of the angular velocity by the moment of inertia $I(\Omega)$ according to
\begin{equation}
J(\Omega) = I(\Omega) \cdot \Omega
\end{equation}
and find 
\begin{equation}
J(\Omega) = \frac{ M \left(  \displaystyle \frac{2GM}{c^{2}} \right)^{2} }
            { 1 +  \left(  \displaystyle \frac{2GM}{c^{2}} \right)^{2} \displaystyle \frac{\Omega^{2}}{c^{2}}   } 
            \cdot \Omega . \\
\end{equation}
In the limit $\Omega \rightarrow 0$, we set $\Omega^{2} = 0$ in the above expression. The rotation stops 
and the Kerr mass becomes the irreducible mass of a Schwarzschild black hole. The angular momentum relation now becomes 
\begin{equation}
J(\Omega) = M_{irr} \left( \frac{2GM_{irr}}{c^{2}} \right)^{2} \cdot \Omega  .     
\end{equation}
A Schwarzschild black hole of mass $M_{S}$ and radius $R_{S}$ is derived and its moment of inertia is remarkably
\begin{equation}
I = M_{S} R_{S}^{2} . \\
\end{equation}

The moment of inertia is an effective way of finding the structure of an unknown body. Equation (10) shows that the
total mass of the Schwarzschild black hole as seen by a distant observer is all located at the Schwarzschild
radius. Its structure is now to be determined. Since the Schwarzschild black hole is known to be perfectly spherical and the External Energy Paradigm states that mass-energy cannot exist inside the horizon, we conclude that the Schwarzschild black hole is effectively a hollow spherical shell in curved space. Note that one cannot calculate the moment of inertia by integrating mass elements as is done in Newtonian mechanics because mass density is not well defined locally in general relativity.\\ 

Thus the hollow massive shell is an intrinsic property of the Kerr metric and it is hidden in the Kerr solution.
The shell would provide a physical surface where electric particles can aggregate instead of simply hovering at the horizon
without cause. The physical shell explains naturally the electrical properties of the membrane as a conductor in the Membrane 
Paradigm. In addition, there is no need to postulate an anti-de Sitter spacetime in the interior of the black hole as is 
proposed in the gravastar model [16] .\\

Once we accept the hollow shell structure of the black hole we find that major paradoxes that have
confounded physicists for the past several decades are suddenly resolved. We could mention the following cases:\\

\noindent
1. {\em The Area Paradox} - It is known that the entropy of a Schwarzschild black hole is porportional to its surface area 
   [17] and not  to its volume. The shell structure naturally explains this since all matter particles are found at the 
   surface and not in the interior. Volume is not a relevant factor in black hole dynamics.\\

\noindent
2. {\em The Singularity Paradox} - Singularity exists inside the black hole because particles are following the geodesics
   in extended Schwarzschild spacetime. In the shell structure, there is no singularity inside the black hole because 
   external particles can never cross the surface.\\

\noindent
3. {\em The Information Loss Paradox} - Information is believed to be lost in Hawking radiation as the infalling
   particles cross the horizon of a black hole [18]. With the shell structure, information carried by quantum particles is 
   only stored at the surface. It may be returned to external spacetime during a rapid process such as the merging of two 
   black holes with huge release of gravitational waves.\\

\noindent   
4. {\em The Firewall Paradox} - It is believed that nothing unusual occurs when particles go through the event horizon 
    according to the Equivalence Principle. However, in the firewall scenario [19], an observer falling into a black hole 
    would encounter strong radiation near the horizon. In our treatment, the physical nature of the shell acts as a firewall 
    for a black hole since infalling particles cannot cross the horizon.\\
   
The mysteries and paradoxes of the black hole seem to have been created by adopting the point mass as the source of gravity and as quantum particles in spacetime. The point mass does not exist in Nature. It is a construction in classical mechanics. The moment of inertia shows that the black hole is not a singularity.\\

\end{document}